\documentclass{vldb}
\usepackage{graphicx}
\usepackage{balance}  
\usepackage{amsmath}
\usepackage{amssymb}
\usepackage{listings, lstautogobble}
\usepackage{framed}
\usepackage{float}
\usepackage{multirow}
\usepackage{url}
\usepackage{times}
\usepackage{courier}
\usepackage{graphicx}
\usepackage{stfloats}
\usepackage{color}
\usepackage{caption}
\usepackage{subcaption}
\usepackage{diagbox}
\usepackage{xspace}
\usepackage{highlighter}
\usepackage[linesnumbered]{algorithm2e}
\SetCommentSty{mycommfont}

\newcommand{\mlopt}{\textsc{TuPAQ}\xspace}
\def\reals{\mathbb{R}}
\SetAlCapSkip{1em}

\begin{document}
%

\title{TuPAQ: An Efficient Planner for Large-scale Predictive Analytic Queries}

%
%
%
%
%

\numberofauthors{5} 
%
\author{
%
%
\alignauthor
Evan R. Sparks\\
       \affaddr{Computer Science Division\\UC Berkeley}\\
       \email{sparks@cs.berkeley.edu}
\alignauthor
Ameet Talwalkar\\
       \affaddr{Computer Science Dept.\\UCLA}\\
       \email{ameet@cs.ucla.edu}
\and  
\alignauthor Michael J. Franklin\\
       \affaddr{Computer Science Division\\UC Berkeley}\\
       \email{franklin@cs.berkeley.edu}
\alignauthor Michael I. Jordan\\
       \affaddr{Computer Science Division\\UC Berkeley}\\
       \email{jordan@cs.berkeley.edu}
\alignauthor 
Tim Kraska\\
       \affaddr{Dept. of Computer Science\\Brown University}\\
       \email{tim\_kraska@brown.edu}
}

\maketitle
\begin{abstract}
The proliferation of massive datasets combined with the development 
of sophisticated analytical techniques have enabled
 a wide variety of novel applications such as improved product
recommendations, automatic image tagging, and improved speech-driven interfaces.
These and many other applications can be supported by Predictive Analytic
Queries (PAQs).
A major obstacle to supporting PAQs is the
challenging and expensive process of identifying and training an appropriate predictive model.
Recent efforts aiming to automate this process have focused on single node
implementations and have assumed that model training itself is a black box, thus
limiting the effectiveness of such approaches on large-scale problems. In this
work, we build upon these recent efforts and propose an integrated PAQ planning
architecture that combines advanced model search techniques, bandit resource
allocation via runtime algorithm introspection, and physical optimization via
batching.
The result is \mlopt, a component of the MLbase system, which solves the PAQ planning problem with comparable quality
to exhaustive strategies but an order of
magnitude more efficiently than the standard baseline approach, and 
can scale to models trained on terabytes of data across hundreds of machines.
\end{abstract}


\section{Introduction}

\label{sec:intro}
Rapidly growing data volumes coupled with the maturity of sophisticated statistical techniques have led to a new type of data-intensive workload: predictive analytics over large scale, distributed datasets. Indeed, the support of predictive analytics is an increasingly active area of database systems research. Several systems that integrate statistical query processing with a data management system have been developed. However, these systems force users to describe their statistical model in dense mathematical notation~\cite{simsql,SystemML} or in terms of a specific model~\cite{MADLib,Zhang:2014tf,Deshpande:2006wi} and provide little guidance about the proper configuration of the model---that is, a user must know that a linear SVM or Kalman filter is a good statistical procedure to answer their query, and configure that procedure appropriately.

\begin{figure}
\begin{subfigure}[b]{0.5\textwidth}
\begin{lstlisting}[language=SQL, frame=shadowbox, breaklines=true, autogobble=true, morekeywords={PREDICT,GIVEN}, escapechar=@, belowskip=\smallskipamount, aboveskip=\smallskipamount]
SELECT vm.sender, vm.arrived,
@\HighlightFrom@PREDICT(vm.text, vm.audio)@\HighlightTo@ 
@\HighlightFrom@GIVEN LabeledVoiceMails@\HighlightTo@
FROM VoiceMails vm 
WHERE vm.user = 'Bob' AND vm.listened is NULL
ORDER BY vm.arrived 
DESC LIMIT 50
\end{lstlisting}
\caption{Speech-to-text transcription.}
\label{fig:vms}
\end{subfigure}
\begin{subfigure}[b]{0.5\textwidth}
\begin{lstlisting}[language=SQL, frame=shadowbox, breaklines=true, autogobble=true, morekeywords={PREDICT,GIVEN}, escapechar=@, aboveskip=\smallskipamount, belowskip=\smallskipamount]
SELECT p.image
FROM Pictures p
WHERE @\HighlightFrom@PREDICT(p.tag, p.photo) = 'Plant' GIVEN@\HighlightTo@ @\HighlightFrom@LabeledPhotos@\HighlightTo@
AND p.likes > 500
\end{lstlisting}
\caption{Photo classification.}
\label{fig:photos}
\end{subfigure}
    \caption{Two examples of PAQs, with the predictive clauses highlighted
in green. (\protect\ref{fig:vms}) returns the predicted text transcription of Bob's voicemails from their audio content. 
(\protect\ref{fig:photos}) finds popular pictures of
photos based on an image classification model---even if the images are not
labeled. Each of these use cases may require considerable training data.}
    \label{fig:paq}
\end{figure}

In our work, our goal is to raise the level of abstraction for data analysts. Instead of choosing a specific statistical model and featurization strategy, we provide a declarative query interface where users declare that they wish to predict an attribute from some other collection of attributes and optionally provide example training data. Given these inputs, the system automatically makes predictions for the target attribute on new data. With our system, users issue Predictive Analytic Queries, or PAQs, which are traditional database queries, augmented with new predictive clauses. Two examples of PAQs are given in Figure~\ref{fig:paq}---with the predictive clauses highlighted. The output of a predictive clause is an attribute like any other---one that can be grouped and sorted on or used in other clauses. The syntax of these predictive clauses is as follows:

\[ 
\text{\texttt{PREDICT} ($a_{predicted}$ [, $a_1$, ..., $a_n$ ]) \texttt{GIVEN} $R$}
\]

Where, $a_{predicted}$ is the attribute to be predicted. $a_1, ..., a_n$ is an optional set of predictor attributes. $R$ is a relation containing training examples with the restriction that $ \left\{a_{predicted}, a_1, ..., a_n\right\} - Attributes(R) = \emptyset $.
This syntax is general enough to support a wide range of predictive tasks---including classification, regression, and item recommendation.
    



Given recent advances in statistical methodology, supervised machine
learning (ML) techniques are a natural way to support the predictive clauses
in PAQs.  In the supervised learning setting, a statistical model is created via
training data to relate the input attributes to the desired output
attribute.  Furthermore, ML methods learn better models as the size of the
training data increases, and recent advances in distributed ML algorithm
development, are aimed at enabling
large-scale model training in the distributed setting.~\cite{agarwal14a,Panda:2009uh,MackeyTaJo11} 

Unfortunately, the
application of supervised learning techniques to a new input dataset is
computationally demanding and technically challenging.
For a non-expert, the process of carefully preprocessing the input
attributes, selecting the appropriate ML model, and tuning its
hyperparameters can be an ad-hoc and time-consuming task. 
For example, to build a predictive model for a classification task like the one shown in Figure~\ref{fig:photos} using conventional tools, a user needs to choose from one of many algorithms for extracting features from image data, then select an appropriate classification model---all the while tuning the configuration of each. 
Finally, the user will need to settle on a strategy to select the best performing model. Failure to follow these steps can lead to models that do not work at all, or worse, provide inaccurate predictions.

In practice, this process of training a supervised model is highly procedural, with even ML experts often having to fall back on standard recipes (e.g. the libSVM guide~\cite{libsvmguide}) in the attempt to to obtain reasonable results.
At scale, the problem of finding a good model is exacerbated, and conventional approaches can be prohibitively expensive. For example, sequential grid search is a popular method implemented in many software packages~\cite{graphlabgrid,scikitlearn,caret}, but as we show, requires many more models to be trained than necessary to achieve good results.

\begin{figure}
  \centering
  \includegraphics[width=1.0\columnwidth]{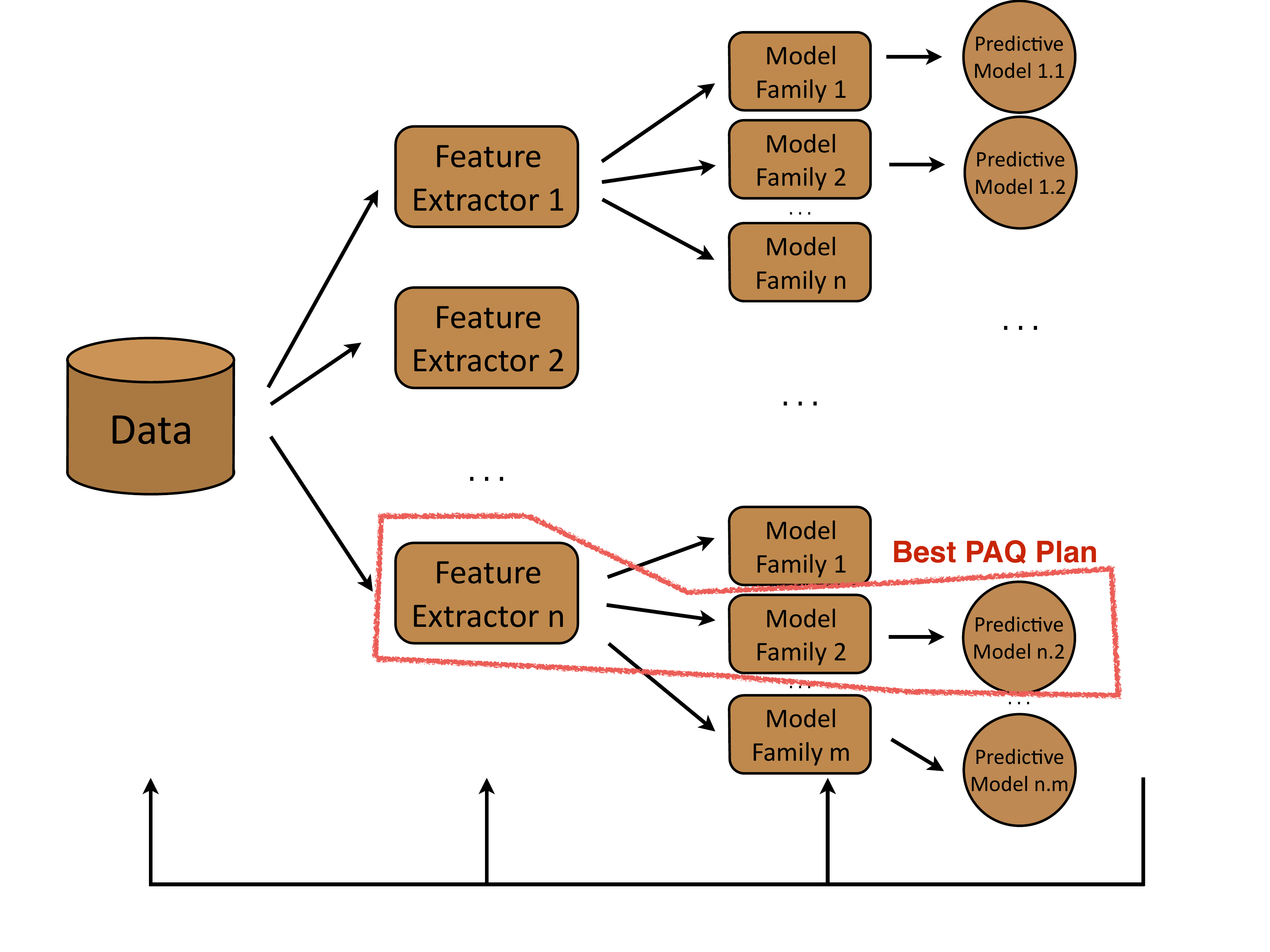}
  \caption{Finding an appropriate predictive model is a process of continuous refinement. Each stage must be carefully tuned to ensure high quality. \mlopt automates this process.}
  \label{fig:pipeline}
\end{figure}

At an abstract level the process of finding a good machine learning model is in some ways analogous to a query planning problem where the challenge is to construct a model from training data given a potential space of model families. 
In the PAQ setting (illustrated in Figure~\ref{fig:pipeline}),
some process must be followed to refine the choice of feature and model selection, and in practice this problem is often iterative. 
The \emph{PAQ planning problem} is the task of efficiently finding a high quality PAQ plan, given training data, a set of candidate statistical model families, and their configurations.

A good PAQ planner will return a high quality PAQ plan efficiently. The \emph{quality}
of a PAQ plan is measured in terms of a statistic relevant to the predictive
task, such as accuracy on validation data in the classification setting or the Mean Squared Error (MSE) in the regression setting.
\emph{Efficiency} is measured in terms of the total time to arrive at such a model---we refer to this as \emph{learning time}. In this work we focus on model search and hyperparameter tuning when the dimensionality of the search space is small, and training budget is also small. This scenario maps well to practical demands, given that there exist a limited set of learning algorithms and hyperparameters for a given predictive learning task, and the cost of training a model may be high, particularly in the large-scale setting. In such scenarios, PAQ planning can lead to substantial improvements in model quality. We further restrict our focus to model family selection and hyperparameter tuning, as opposed to also considering selection of appropriate featurizers. However, we believe the techniques presented here can be generalized to more complicated predictive analytic pipelines, since the number of established featurization techniques for a given domain is also limited.

In the remainder of the paper, we explore the challenges associated with PAQ planning
and introduce a PAQ planner for the MLbase system~\cite{Kraska13} for declarative machine learning, called the \textbf{T}raining-s\textbf{u}pported
\textbf{P}redictive \textbf{A}nalytic \textbf{Q}uery planner (\mlopt). The goal of \mlopt in the context of MLbase is to tackle the PAQ planning problem at scale. Using advanced search techniques, bandit resource allocation, and batching optimizations, \mlopt identifies a suitable model to satisfy a
user's high-level declarative query, thus differing from  previous systems~\cite{MADLib,
Deshpande:2006wi, Wang:2008vy, Zhang:2014tf}
focused on predictive analytics, which force users to perform model
search themselves.  Further, although the ML community has developed techniques
for finding good configurations for learning models, none have focused on
applying these techniques in the large-scale setting.  


With \mlopt, we make the following contributions: 
\begin{itemize}
  \item We introduce PAQs, a declarative query interface that
  enables analysts to operate on imputed attribute values. 
  \item We demonstrate the effectiveness of supervised ML techniques at
  supporting PAQs, both in terms of high accuracy and efficiency, especially when compared with basic approaches.
  \item We describe the \mlopt algorithm for PAQ planning which combines
  logical optimization via model search and physical
  optimization via batching and bandit resource allocation via runtime
  introspection. 
  \item We describe an implementation of the \mlopt algorithm in Apache Spark, building on our earlier work on the MLbase architecture~\cite{Kraska13}. 
  \item We evaluate several points in the design space with respect to each
  of our logical and physical optimizations, and demonstrate that proper selection of each can
  dramatically improve both accuracy and efficiency. 
  \item We present experimental results on large, distributed datasets up to
terabytes in size, demonstrating that \mlopt converges to high quality PAQ
plans an order of magnitude faster than a simple PAQ planning strategy.
\end{itemize}

The remainder of this paper is organized as follows.
Section~\ref{sec:paq_planning} formally defines the PAQ planning problem, explains its connection to traditional database query optimization research, and introduces a standard baseline approach for the problem, and provides a
high-level overview of \mlopt. We next present details about \mlopt's three
main components in Section~\ref{sec:optimizations}, highlighting the design decisions
for each of these components.  Section~\ref{sec:designeval} subsequently presents an
empirical study of this design space and a comparison with traditional methods for solving this problem. We then present results on a large-scale
evaluation of \mlopt in Section~\ref{sec:evaluation}, with each of \mlopt's three
components tuned based on the results of the previous section. 
In Section~\ref{sec:related} we explore the relationship between PAQ planning and existing works related to supporting predictive and analytical workloads.
We conclude with Section~\ref{sec:futureworkconclusions}, which summarizes our work and
discusses future extensions to \mlopt. 

\section{PAQ Planning and T{\normalsize{\textbf U}}PAQ}
\label{sec:paq_planning}
In this section, we define the PAQ planning problem in more detail, and describe its relationship to traditional query optimization. Then, we discuss two approaches to PAQ planning. 
The first, which we call the baseline approach, is inspired by common practice.
The second approach, \mlopt, allows us to take advantage of logical and physical optimizations in the planning process.
\mlopt has a rich design space, which we describe in further detail in Section~\ref{sec:optimizations}.
Finally, we describe how \mlopt fits into the broader MLbase architecture.

\subsection{Defining PAQ Planning}
Figure~\ref{fig:paq} shows several example PAQs that, in practice, need extremely large training datasets with millions of examples each with hundreds of thousands of features to return accurate results.
Other PAQs that need large training sets include problems in image classification,
speech-to-text translation, and web-scale text mining.
As defined in Section~\ref{sec:intro}, PAQs can be any query where an attribute or predicate value must be imputed to complete the query. In this work we concern ourselves specifically with PAQs that can be answered based on user-supplied labeled training data, typically of the same format as the data for which values are to be imputed.
We focus specifically on the components of the system that are necessary to efficiently support clauses of the form shown in Section~\ref{sec:intro}.
While the strategies discussed here can operate in situations where queries have joined relations or complex aggregates, we expect that future work will explore optimizations specific to these situations.

The PAQ planner's job is to find a PAQ plan that maximizes
some measure of quality (e.g., in terms of goodness of fit to held-out data) in a short amount of time, where learning time is constrained by
some budget in terms of the number of models considered, total
execution time, or the number of scans over the training data.
The planner thus takes as input a training dataset, a description of a space
of models to search, and some budget or stopping criterion.  The description of
the space of models to search includes the set of model families to search over
(e.g., SVM, decision tree, etc.) and reasonable ranges for their associated
hyperparameters (e.g., regularization parameter for a regularized linear model
or maximum depth of a decision tree).  The output of a PAQ Planner is a plan  that can be applied to unlabeled
data points to obtain a prediction for the desired attribute. In the context of \mlopt, this plan is a statistical model that can be applied to unseen training data.

In this work, we operate in a scenario where individual models are of
dimensionality $d$, where $d$ is less than the total number of
example data points $N$. Note that $d$ can nonetheless be quite large, e.g.,
$d=200,000$ in our large-scale speech experiments and $d=160,000$ in our large scale image experiments (see Section~\ref{sec:evaluation}). Recall that in this paper, we focus on classification, and consider a small number of model families, $f \in F$,
each with several hyperparameters, $\lambda \in \Lambda$.
These assumptions map well to reality, as there are a handful of
general-purpose classification methods that are deployed in practice. Further, we expect that these techniques will naturally apply to other supervised learning tasks---such as regression and collaborative filtering, which may only differ in terms of their definition of plan quality.
We evaluate the quality of each plan by computing accuracy on
held-out datasets, and we measure learning time as the amount of time required
to explore a fixed number of models from some model space. In our large-scale
distributed experiments (see Section~\ref{sec:evaluation}) we report parallel run times. 

Additionally, in this paper we focus on model families that are trained via multiple sequential scans of the training data.  In particular, we focus
on three model families: linear Support Vector Machines (SVM), logistic regression trained via gradient descent, and nonlinear
SVMs using random features~\cite{RandomKitchenSink} trained via block coordinate descent. 
This iterative sequential access pattern encompasses a wide
range of learning algorithms, especially in the large-scale distributed
setting.  For instance, efficient distributed implementations of linear
regression~\cite{Franklin13}, tree based models~\cite{Panda:2009uh}, Naive
Bayes classifiers~\cite{Franklin13}, and $k$-means clustering~\cite{Franklin13}
all follow this same access pattern.

\subsection{Connections to Query Optimization}
Given that PAQ planning is the automation of a declaratively specified task, it is natural to draw connections to decades worth of relational query optimization research when
tackling the PAQ planning problem.  Traditional database systems invest in
the costly process of query planning to determine a good execution plan that can
be reused repeatedly upon subsequent execution of similar queries.
While query planning for a PAQ involves the costly process of
identifying a high quality predictive model, this cost is offset by the subsequent
ability to perform near real-time PAQ evaluation.  Additionally, both types
of query planning can be viewed as search problems, with traditional query
planning searching over the space of join orderings and access methods, and PAQ planning
searching over the space of machine learning models.
 
There are some notable differences between these two problems, however, leading
to a novel set of challenges to address in the context of PAQ planning.
First, unlike traditional database queries, PAQs do not have unique answers
due to the inherent uncertainty in predictive models learned from finite
datasets.  Hence, PAQ planning focuses on both quality and efficiency
(compared to just efficiency for traditional query planning), and needs to
balance between these goals when they conflict.  Second, the search space
for PAQs is not endowed with well-defined algebraic properties, as it
consists of possibly unrelated model families and feature extractors, each
with its own access patterns and hyperparameters.  Third, evaluating a
candidate query plan is expensive and in this context involves learning
the parameters of a statistical model.  Learning the parameters of a
single model can involve upwards of hundreds of passes over the input
data, and there exist few heuristics to estimate the effectiveness of a
model before this costly training process.

Now, we turn our attention to algorithms for PAQ planning.

\subsection{Baseline PAQ Planning}



The conventional
approach to PAQ planning is sequential grid search~\cite{graphlabgrid,scikitlearn,caret}. For instance, consider the
tag prediction example in Figure~\ref{fig:photos}, in which the PAQ is processed
via an underlying classification model trained on \texttt{LabeledPhotos}.
Moreover, consider a single ML model family for binary
classification---logistic regression---which has two hyperparameters: learning
rate and regularization.  Sequential grid search divides the
hyperparameter space into a grid and iteratively trains models at these grid
points.


Grid search has several shortcomings.  First, the results of previous
iterations in the sequence of grid points are not used to inform future iterations of search.  Second, the
curse of dimensionality limits the usefulness of this method in high
dimensional hyperparameter spaces. Third, grid points may not represent a good
approximation of global minima---true global minima may be hidden between grid
points, particularly in the case of a very coarse grid.  Nonetheless, sequential grid search 
is commonly used in practice, and is a natural baseline for PAQ planners.

In Algorithm~\ref{alg:naiveplanner}, we show the logic encapsulated in such a baseline PAQ planner.
In this example, the budget is the total number of models to train.

	\begin{algorithm}[h]
	 \SetKwInOut{Input}{input}\SetKwInOut{Output}{output} 
	 \Input{LabeledData, ModelSpace, Budget}
	 \Output{BestModel}	 
	 bestModel $\gets$ $\emptyset$\;
	 grid $\gets$ gridPoints(ModelSpace, Budget)\;
	 \While{Budget $>$ 0}{
	   proposal $\gets$ nextPoint(grid)\;
	   model $\gets$ train(proposal, LabeledData)\;
	   \If{quality(model) $>$ quality(bestModel)}
	     {bestModel $\gets$ model\;}
	   Budget $\gets$ Budget $-$ 1\;}
	 \Return{bestModel};
	 
	 \caption{A baseline PAQ planning procedure with conventional grid search. The function ``gridPoints'' returns a coarse grid over the dimensions of model space, where the total number of grid points is determined by the budget.}
	 \label{alg:naiveplanner}
	\end{algorithm}

\subsection{T{\normalsize{\textbf U}}PAQ Planning}
As discussed in the previous section, grid search is a suboptimal search method
despite its popularity.  Moreover, from a systems perspective, the algorithm illustrated in
Algorithm~\ref{alg:naiveplanner} 
has additional drawbacks beyond those of grid search. In
particular, this procedure performs sequential model training and also treats
the training of each model as a black-box procedure.

\begin{algorithm}[]
 \SetKwInOut{Input}{input}\SetKwInOut{Output}{output} 
 \Input{LabeledData, ModelSpace, Budget, PartialIters, BatchSize}
 \Output{BestModel}
 bestModel $\gets$ $\emptyset$\;
 history $\gets$ []\;
 proposals $\gets$ []\;
 freeSlots $\gets$ batchSize\;

 \While{Budget $>$ 0}{
   freeSlots $\gets$ batchSize - length(proposals)\;
   proposals $\gets$ proposals + proposeModels(freeSlots, ModelSpace, history)\tcp*{Model Search}
   models $\gets$ trainPartial(proposals, LabeledData, PartialIters)\tcp*{Batching}
   Budget $\gets$ Budget $-$ len(models)$*$PartialIters\;
   
   (finishedModels, history, proposals) $\gets$ banditAllocation(models, history)\tcp*{Bandits}
   
   \For{m in finishedModels}{
     \If{quality(m) $>$ quality(bestModel)}{
       bestModel $\gets$ m\;
     }
   }
 }
 \Return(bestModel);
  
 \caption{The planning procedure used by \mlopt.}
 \label{alg:mloptplanner}
\end{algorithm}

In contrast, we propose the \mlopt algorithm, described in Algorithm~\ref{alg:mloptplanner}, to
address all three of these shortcomings via logical and physical optimizations.
First, the \mlopt algorithm allows for more \emph{sophisticated search strategies}.
Line 7 shows that our model search procedure can now use training history as
input. Here, ``proposeModel'' can be an arbitrary model search procedure.
Second, our algorithm performs \emph{batching} to train multiple models
simultaneously (Line 8). Third, our algorithm deploys \emph{bandit resource allocation via
runtime inspection} to make on-the-fly decisions. Specifically, the algorithm
compares the quality of the models currently being trained with historical
information about the training process, and determines which of the current
models should be trained further (Line 10).  

These three optimizations are discussed in detail in Section~\ref{sec:optimizations}, with a focus on the design space for each of them.
In Section~\ref{sec:designeval} we then evaluate the options in this design
space experimentally, and then in Section~\ref{sec:evaluation} compare the
baseline algorithm (Algorithm \ref{alg:naiveplanner}) to \mlopt running with
good choices for search method, batch size, and bandit allocation criterion,
i.e., choices informed by the results of Section~\ref{sec:designeval}.

Before exploring the design space for the \mlopt algorithm, we first describe how \mlopt fits into a larger system to support PAQs.

\begin{figure}
  \centering
  \includegraphics[width=0.8\columnwidth]{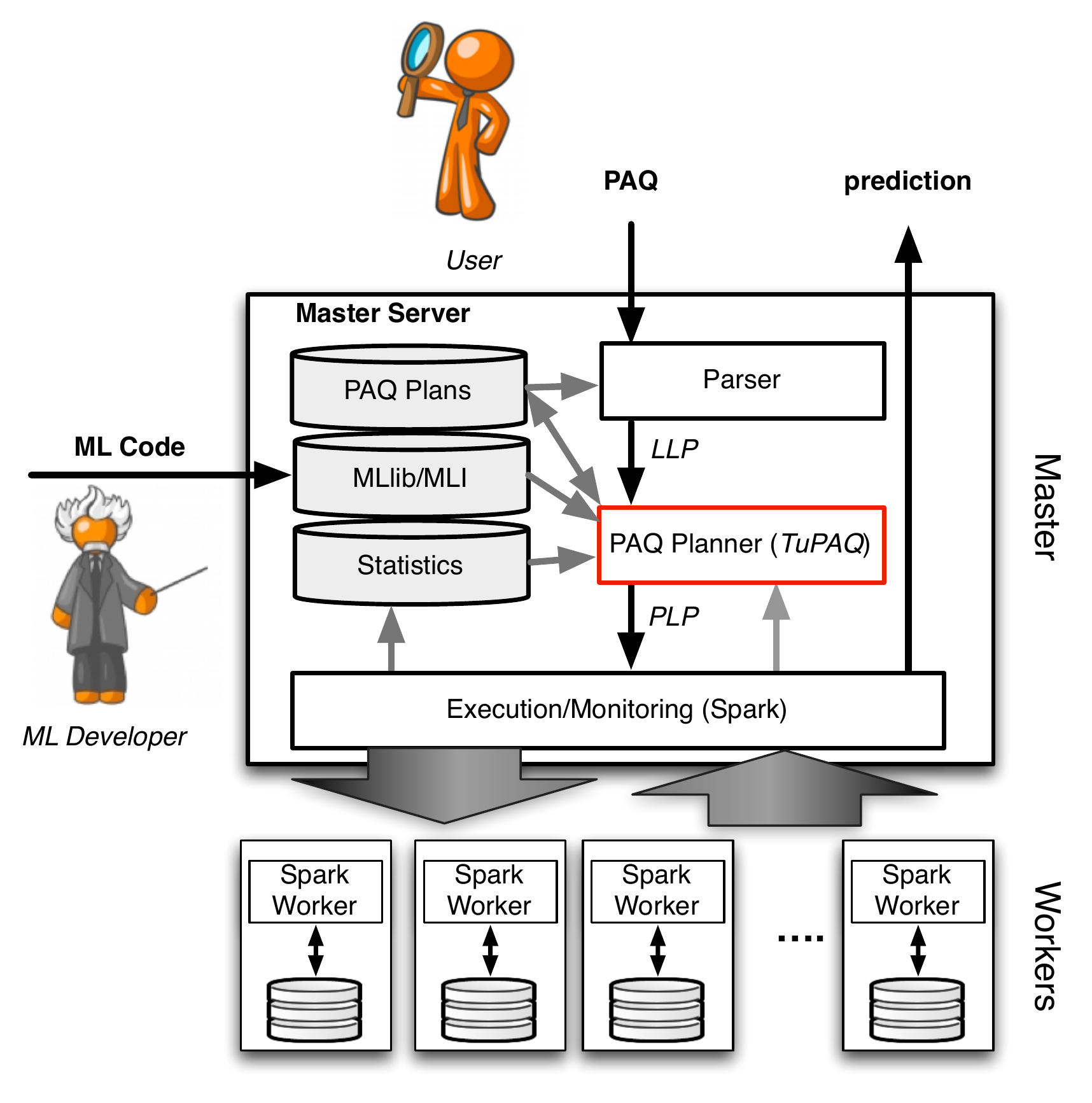}
  \caption{The \mlopt planner is a critical component of the MLbase~\protect\cite{Kraska13} system for simplified large scale machine learning. \mlopt interacts with a distributed run-time and existing machine learning algorithms to efficiently find a PAQ plan which yields high quality predictions.}
  \label{fig:mlbarch}
\end{figure}

\subsection{T{\normalsize{\textbf U}}PAQ and the MLbase Architecture}

\mlopt lies at the heart of MLbase~\cite{Kraska13}, a novel system designed to simplify the process of implementing and using scalable machine learning techniques.
By giving users a declarative interface for machine learning tasks, the problem of hyperparameter tuning and feature selection can be pushed down into the system. The architecture of this system is shown in Figure~\ref{fig:mlbarch}.

At the center of the system, some optimizer or planner must be able to quickly identify a suitable model for supporting predictive queries. 
We note that the system described in this paper introduces some key architectural differences compared with
the original MLbase architecture in \cite{Kraska13}.
In particular, we make the concept of a ``PAQ planner'' explicit, and introduce a catalog for PAQ plans. 
When a new PAQ arrives, it is passed to the planner which determines whether a new PAQ plan needs to be created.
The entire system is built upon Apache Spark, a cluster compute system designed for iterative computing~\cite{Spark}, and we leverage MLlib, and other components present in Apache Spark, as well as MLI~\cite{MLI}.



\section{T{\normalsize{\textbf U}}PAQ Design Choices}
\label{sec:optimizations}
 


In this section, we examine the design choices available to the \mlopt planner.
As stated previously, we are targeting algorithms
that run on tens to thousands of nodes in commodity computing clusters, and training datasets
that fit comfortably into cluster memory---on the order of tens of gigabytes to
terabytes.  Training of a single model to convergence on such a cluster is expected to
require tens to hundreds of passes through the training data, and may take on the order of minutes.
Moreover, with a multi-terabyte dataset, performing a sequential grid search
involving even just 100 model configurations each with a budget of 100 scans of the training data could take hours to days of processing time, even assuming that the algorithm runs at memory speed.
Hence, in this regime the baseline PAQ planner is tremendously costly.

Given the design choices presented in Section~\ref{sec:paq_planning}, we ask how these design choices might be optimized to provide fast, high quality PAQ planning.
In the remainder of this
section we present the following optimizations---advanced model search techniques, bandit resource allocation via
runtime algorithm introspection, and physical optimization via
batching---that in concert provide \mlopt with an order-of-magnitude gain in performance.

\subsection{Better Model Search}
We call the problem of finding the highest quality model from a space
of model families and their hyperparameters the \emph{model search} problem, and the solution to this problem is of central importance to \mlopt.
We view model search as an optimization problem over a potentially
non-smooth, non-convex function in high dimensional space, where it is
expensive to evaluate the function and for which we have no closed form
expression for the function to be optimized (and hence cannot compute
derivatives).  Although grid search remains the standard solution to this problem, various alternatives have been proposed for the general problem
of derivative-free optimization, some of which are particularly tailored for
the model search problem.  Each of these methods provides an opportunity to
speed up \mlopt's planning time, and in this section we provide a brief survey
of the most commonly used methods.  In Section~\ref{sec:designeval} we
evaluate each method on several datasets to determine which method is most
suitable for PAQ planning.

Traditional methods for derivative-free optimization include grid search (the
baseline choice for a PAQ planner) as well as random search, Powell's method
\cite{powell1964efficient}, and the Nelder-Mead method \cite{Nelder:1965tk}.  
Given a hyperparameter space, grid search
selects evenly spaced points (in linear or log space) from this space, while
random search samples points uniformly at random from this space.
Powell's method can be seen as a derivative-free analog to coordinate
descent, while 
the Nelder-Mead method can be roughly interpreted as a derivative-free analog to
gradient descent.

Both Powell's method and the Nelder-Mead method expect unconstrained search
spaces, but function evaluations can be modified to severely penalize exploring
out of the search space.  However, both methods require some degree of
smoothness in the hyperparameter space to work well, and can easily get stuck
in local minima.  Additionally, neither method lends itself well to categorical
hyperparameters, since the function space is modeled as continuous.  For these
reasons, we are unsurprised that they are inappropriate methods to use in the
model search problem where optimization is done over an unknown function that
is likely non-smooth and not convex.

More recently, various methods specifically for model search have been recently
introduced in the ML community, including Tree-based Parzen Estimators
(HyperOpt)~\cite{Bergstra:2011tj}, Sequential Model-based Algorithm Configuration (Auto-WEKA)~\cite{Thornton:2013ea} and Gaussian Process based methods, e.g.,
Spearmint~\cite{Snoek:2012vl}. These algorithms all share the property that they can
search over spaces which are nested (e.g. multiple model families) and accept
categorical hyperparameters (e.g. regularization method). HyperOpt begins with a
random search and then probabilistically samples from points with more promising
minima, Auto-WEKA builds a Random Forest model from observed hyperparameter results,
and  Spearmint implements a Bayesian methods based on Gaussian Processes.  

\subsection{Bandit Resource Allocation}
\label{ssec:bandit}

Models are not all created equal. In the context of model
search, typically only a fraction of the models are of high-quality,
with many of the
remaining models performing drastically worse.  Under certain
assumptions, allocating resources among different model configurations can be
naturally framed as a multi-armed bandit problem~\cite{BubeckC12}. Indeed, assume we are given a
\emph{fixed set} of $k$ model configurations to evaluate, as in the case of
grid or random search, along with a fixed budget $B$.  Then, each model can be
viewed as an `arm' and the model search problem can be cast as a $k$-armed
bandit problem with $T$ rounds.  At each round we perform a single iteration of
a particular model configuration, and return a reward indicating
the quality of the updated model, e.g., validation accuracy.  In such settings,
multi-armed bandit algorithms can be used to determine a scheduling policy to
efficiently allocate resources across the $k$ model configurations.  Typically,
these algorithms keep a running score for each of the $k$ arms, and at each
iteration choose an arm as a function of the current scores.

\begin{algorithm}[]
 \SetKwInOut{Input}{input}\SetKwInOut{Output}{output} 
 \Input{Models, History}
 \Output{FinishedModels, History, Proposals}
  Proposal $\gets$ []\; 
  FinishedModels $\gets$ []\; 
  bestModel = getBestFromHistory(History)\; 
  \For{m in models}{
     history.append(m)\;
     \uIf{fullyTrained(m)}{
       FinishedModels.append(m)\;
     }
     \ElseIf{quality(m) $ * (1+\epsilon) > $ quality(bestModel)}{
       proposals.append(m)\;
     }
  }
 \Return(FinishedModels, History, Proposals);

 \caption{The bandit allocation strategy used by \mlopt.}
 \label{alg:bandit}
\end{algorithm}

Our setting differs from this standard setting in two crucial ways.  First,
several of our search algorithms select model configurations to evaluate in an
iterative fashion, so we do not have advanced access to a fixed set of $k$
model configurations. Second, in addition to efficiently allocating resources,
we aim to return a reasonable result to a user as quickly as possible, and
hence there is a benefit to finish training promising model configurations once
they have been identified. 

Our bandit selection strategy is a variant of the action elimination algorithm of \cite{EvenDar:2006tf}, and to our knowledge this is the first time this algorithm has been applied to hyperparameter tuning.  Our strategy is
detailed in Algorithm~\ref{alg:bandit}.  This strategy preemptively prunes
models that fail to show promise of converging.  For each model
(or batch of models), we first allocate a fixed number of iterations for
training; in Algorithm~\ref{alg:mloptplanner} the trainPartial() function
trains each model for PartialIters iterations.  Partially trained models are
fed into the bandit allocation algorithm, which determines whether to train the
model to completion by comparing the quality of these models to the quality of
the best model that has been trained to date.  Moreover, this comparison is
performed using a slack factor of $(1+\epsilon)$; in our experiments we set
$\epsilon = .5$ and thus continue to train all models with quality 
within $50\%$ of the best quality model observed so far.
The algorithm stops allocating further resources to models that fail this test,
as well as to models that have already been trained to completion.

\subsection{Batching}

Batching is a
natural system optimization in the context of training machine learning models, with applications
for cross validation and ensembling~\cite{Kumar:2013vm,Canny:2013th}. For PAQ planning,
we note that the access pattern over the training set is identical for many machine learning algorithms.
Specifically, each algorithm takes multiple passes over the input data and updates some intermediate state
(e.g., model weights) during each pass.
As a result, it is possible to batch together the training of multiple models effectively sharing scans across multiple model estimations. 
In a data parallel distributed environment, this has several advantages:
\begin{enumerate}
  \item Better CPU utilization by reducing wasted cycles.
	\item Amortized task launching overhead across several models at once.
	\item Amortized network latency across several models at once.
\end{enumerate}
Ultimately, these three advantages lead to a significant reduction in learning time.
We take advantage of this optimization in line 8 of Algorithm~\ref{alg:mloptplanner}.

For concreteness and simplicity, we will focus on one algorithm---logistic regression trained via gradient descent---for the remainder of this section, but we note that these techniques apply to many model families and learning algorithms.

\subsubsection{Logistic Regression}
Logistic Regression is a widely used machine learning model for binary classification.
The procedure estimates a set of model parameters, $w \in \reals^d$, given a set of data features $X \in \reals^{n \times d}$, and binary labels $y \in {0,1}^n$.
The optimal model $w^*\in\reals^d$ can be found by minimizing the negative likelihood function, $f(w) = -\log p(X|w)$.
Taking the gradient of the negative log likelihood, we have:
\begin{equation}
\label{eq:grad}
\nabla f = \sum_{i=1}^n \bigg [\big (\sigma(w^\top x_i)
- y_i \big ) x_i \bigg ]\,,
\end{equation}
where $\sigma$ is the logistic function.
The gradient descent algorithm (Algorithm~\ref{fig:logreg}) must evaluate this gradient function for all input data points, a task which can be easily performed in a data parallel fashion. Similarly, minibatch Stochastic Gradient Descent (SGD) has an identical access pattern and can be optimized in the same way by working with contiguous subsets of the input data on each partition.

\begin{algorithm}[]
 \SetKwInOut{Input}{input}\SetKwInOut{Output}{output} 
 \Input{X, LearningRate, MaxIterations}
 \Output{Model}
 $i \gets 0$\;
 Initialize Model\;
 \While{$i < MaxIterations$}{
  read current\;
  Model $\gets$ Model - LearningRate * Gradient(Model, X)\;
  $i \gets i + 1$\;
 }
 \caption{Pseudocode for convex optimization via gradient descent.}
 \label{fig:logreg}
\end{algorithm}

The above formulation represents the computation of the gradient by taking a single point and single model at a time.
We can naturally extend this to multiple models simultaneously if we represent our models as a matrix $W \in \reals^{d \times k}$, where $k$ is the number of models we want to train simultaneously, i.e.,

\begin{equation}
\label{eq:batchgrad}
\nabla f = \bigg [X^\top \big (\sigma(X W)
- y \big ) \bigg ]\,.
\end{equation}

This operation can be easily parallelized across data items with each worker in a distributed system computing the portion of the gradient for the data that it stores locally.
Specifically, the portion of the gradient that is derived from the set of local data is computed independently at each machine, and these gradients are simply summed at the end of an iteration. 
The size of the partial gradients (in this case $O(d \times k)$) is much smaller than the actual data (which is $O(n \times d)$), so overheads of transferring these over the network is relatively small.
For large datasets, the time spent performing this operation is almost completely determined by the cost of performing two matrix multiplications---the input to the $\sigma$ function which takes $O(ndk)$ operations and requires a scan of the input data as well as the final multiply by $X^\top$ which also takes $O(ndk)$ operations and requires a scan of the data. 
This formulation allows us to leverage high performance linear algebra libraries that implement BLAS~\cite{blas}---these libraries are tailored to execute exactly dense linear algebra operations as efficiently as possible and are automatically tuned to the architecture we are running on via \cite{atlas}.

The careful reader will note that if individual data points are of sufficiently low dimension, the gradient function in Equation~\ref{eq:grad} can be executed in a single pass over the data from main memory because the second reference to $x_i$ will likely be a cache hit, whereas we assume that $X$ is big enough that it is unlikely to fit entirely in CPU cache.
We examine this effect more carefully in Section~\ref{sec:evaluation}.

\subsubsection{Machine Balance}
One obvious question the reader may ask is why implementing these algorithms via matrix-multiplication should offer speedup over vector/vector versions of the algorithms.
After all, the runtime complexities of both algorithms are identical.
However, modern x86 machines have been shown to have processor cores that significantly outperform their ability to read data from main memory~\cite{McCalpin1995}.  
In particular, on a typical x86 machine, the hardware is capable of reading 0.45B doubles/s from main memory per core, while the hardware is capable of executing 6.8B FLOPS in the same amount of time~\cite{McCalpin2007}.
Specifically, on the machines we tested (Amazon \texttt{c3.8xlarge} EC2 instances), LINPACK reported peak GFLOPS of 110 GFLOPS/s when running on all cores, while the STREAM benchmark reported 60GB/s of throughput across 16 physical cores. This equates to a machine balance of approximately 15 FLOPS per double precision floating point number read from main memory if the machine is using both all available FLOPs and all available memory bandwidth solely for its core computation.
This approximate value for the machine balance suggests an opportunity for optimization by reducing unused resources, i.e., wasted cycles.
By performing more computation for every number read from memory, we can reduce this resource gap.

The Roofline model~\cite{Williams:2009cx} offers a more formal way to study this effect.
According to the model, total throughput of an algorithm is bounded by the smaller of 1) peak floating point performance of the machine, and 2) memory bandwidth times operational intensity of the algorithm, where operational intensity is a function of the number of FLOPs performed per byte read from memory.
That is, for an efficiently implemented algorithm, the bottleneck is either I/O bandwidth from memory or CPU FLOPs. 

Analysis of the unbatched gradient descent algorithm reveals that the number of FLOPs required per byte is quite small---just over 2 flops per number read from memory---a multiply and an add---and since we represent our data as double-precision floating point numbers, this equates to 1/2 FLOP per byte. 
Batching allows us to move ``up the roofline'' by increasing algorithmic complexity by a factor of $k$, our batch size.
The exact setting of $k$ that achieves balance (and maximizes throughput) is hardware dependent, but we show in Section~\ref{sec:evaluation} that on modern machines, $k=10$ is a reasonable choice.

\subsubsection{Amortized Overheads}
In the context of a distributed machine learning system like MLbase, which runs on Apache Spark, delays due to task scheduling and serialization/deserialization can be significant relative to the time spent computing. 

By batching our updates into tasks that require more computation, we are able to reduce the aggregate overhead of launching new tasks substantially.
Assuming a fixed scheduling delay of 200ms per task, if we have a batch size of $k=10$ models, the average task overhead per model iteration drops to 20ms. 
Over the course of hundreds of iterations for hundreds of models, the savings can be substantial.

For a typical distributed system, a model update requires at least two network round-trips to complete: one to launch a task on each worker node, and one to report the results of the task back to the master.
If we were indeed bound by network messaging, then amortizing this cost across multiple models could substantially reduce total overhead due to network latency.
In our experiments, however, the number of messages is relatively small and the network overheads are substantially lower than scheduling and computation overheads, so future work in this setting should focus on minimizing scheduling overheads.








\section{Design Space Evaluation}
\label{sec:designeval}
Now that we have laid out the possible optimizations available to \mlopt, we identify available speedups from these optimizations.
We validated the strategies for model search and bandit resource allocation on five representative datasets across a variable sized model fitting budget.
In particular, these results motivated which model search strategies to incorporate into \mlopt and validate our bandit approach.
Next, we tested our batching optimizations on datasets of various sizes in a cluster environment, to better understand the impact of batching as models and datasets get bigger.

In all experiments, before training, we split our base datasets into $70\%$ training, $20\%$ validation, and $10\%$ testing.
In all cases, models are fit to minimize classification error on the training set, while model search occurs based on classification error on the validation set (validation error).\footnote{While we have thus far discussed model quality, for the remainder of the paper we report validation error, i.e., the inverse of quality, because it is more commonly reported in practice.}
We only report validation error numbers here, but test error was similar.
\mlopt is capable of optimizing for arbitrary performance metrics as long as they can be computed mid-flight, and extends to other supervised learning scenarios.

\subsection{Model Search}
We validated our multiple model search strategies on a series of small datasets with well-formed binary classification problems embedded in them.
These datasets come from the UCI Machine Learning Repository~\cite{ucirepo}.
The model search task involved tuning four hyperparameters---learning rate, L2 regularization parameter, size of a random projection matrix, and noise associated with the random feature matrix.
The random features are constructed according to the procedure outlined in \cite{RandomKitchenSink}.
To accommodate for the linear scale-up that comes with adding random features,
we down sample the number of data points for each model training by the same
proportion.

Our ranges for these hyperparameters were learning rate $\in (10^{-3}, 10^{1})$, regularization $\in (10^{-4}, 10^{2})$, projection size $\in (1 \times d, 10 \times d)$, and noise $\in (10^{-4}, 10^{2})$.

\begin{figure*}[ht]
\centering
\includegraphics[width=0.7\textwidth]{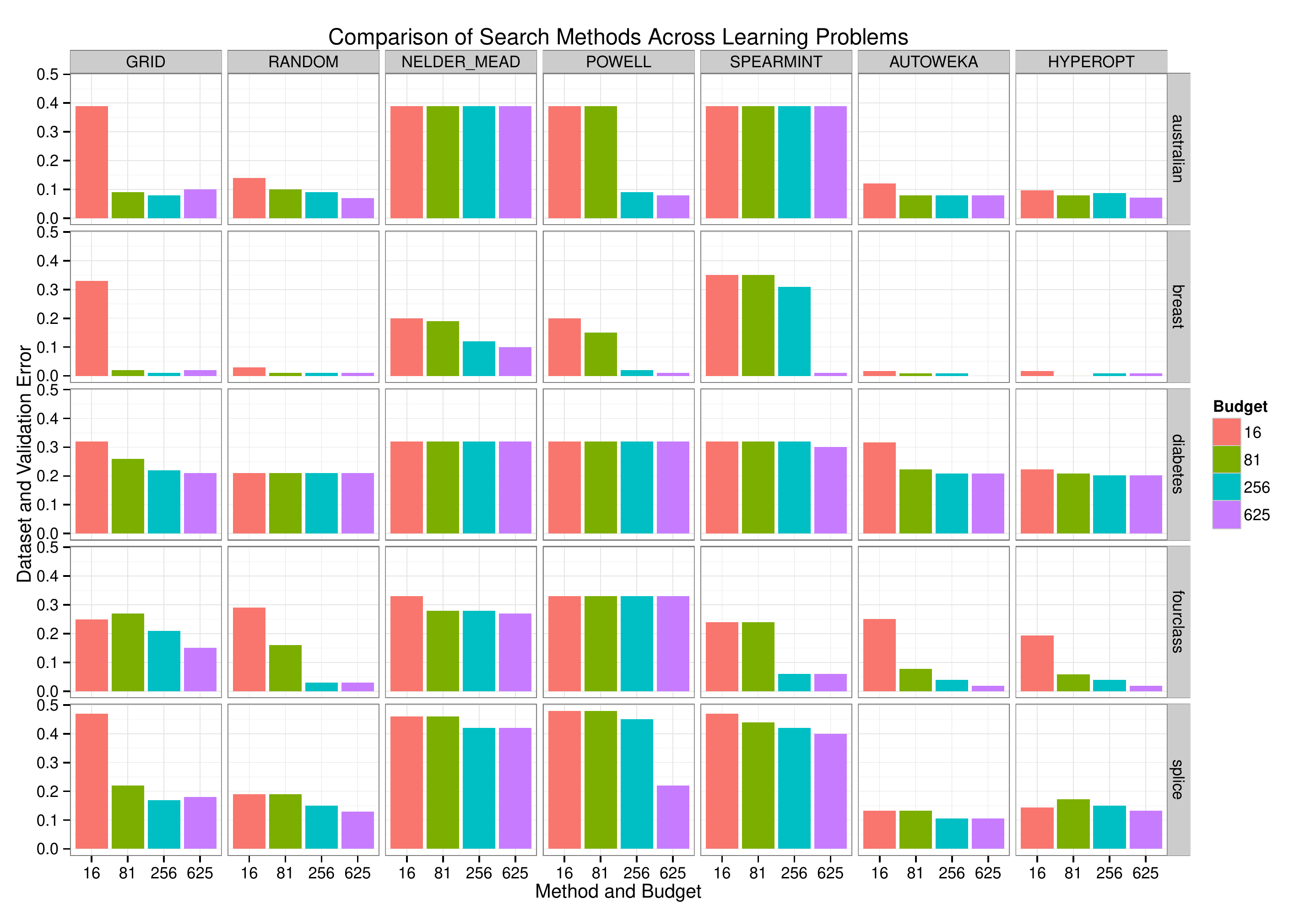}
\caption{Search methods were compared across several datasets with a variable number of function evaluations. Classification error on a validation dataset is shown for each combination. HyperOpt and Auto-WEKA provide state of the art results, while random search performs best of the classic methods.}
\label{fig:searchexps}
\end{figure*}

We evaluated seven search methods: grid search, random search, Powell's method, the Nelder-Mead method, Auto-WEKA, HyperOpt, and Spearmint. 

Each dataset was processed with each search method with a varying number of function calls, chosen to align well with a regular grid of $n^4$ points where we vary $n$ from 2 to 5. 
This restriction on a regular grid is only necessary for grid search but included for comparability.

Results of the search experiments are presented in Figure~\ref{fig:searchexps}. 
Each tile represents a different dataset/search method combination. 
Each bar within the tile represents a different budget in terms of function calls/models trained.
The height of each bar represents classification error on the validation dataset.

With this experiment, we are looking for methods that converge to good models in as small a budget as possible.
Of all methods tried, HyperOpt and Auto-WEKA tend to achieve this criteria best, but random search is not far behind.
We chose to integrate HyperOpt into the larger experiments because it performed
slightly better than Auto-WEKA.
Our architecture fully supports additional search methods, and we expect to implement additional methods in our system over time.

\subsection{Bandit Resource Allocation}
We evaluated the \mlopt bandit resource allocation scheme on the same datasets with random search and 625 total function evaluations---the same as the maximum budget in the search experiments.
The key question to answer here was whether we could identify and terminate poorly performing models early in the training process without significantly affecting overall search quality. 

\begin{figure}
\centering
\includegraphics[width=\columnwidth]{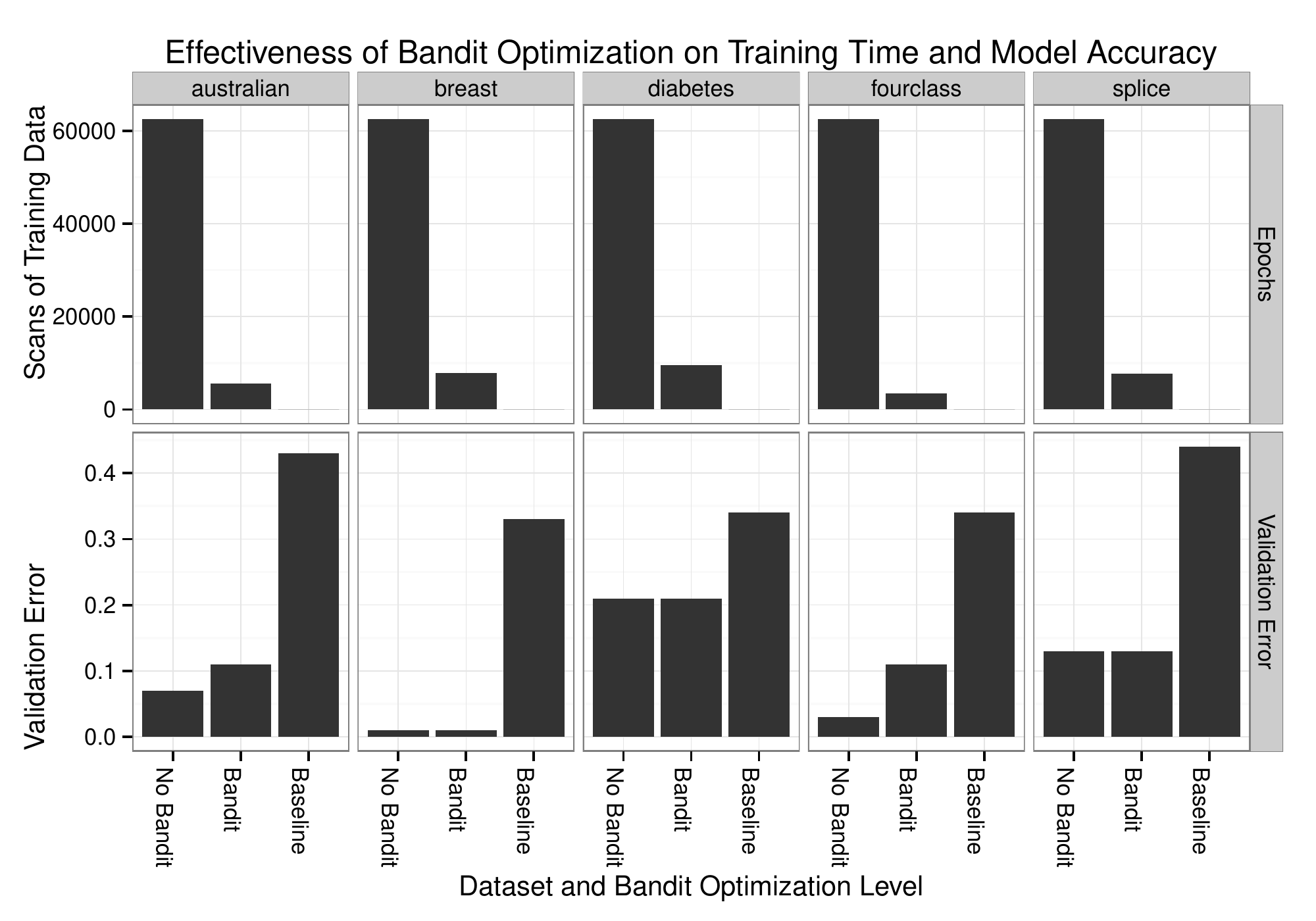}
\caption{Here we show the effects of bandit resource allocation on trained model performance. Model search completes in an average of 83\% fewer passes over the training data than without bandit allocation. Except in one case, validation error is nearly indistinguishable vs. the case where we do not employ the bandit strategy.}
\label{fig:earlystopexps}
\end{figure}

In Figure~\ref{fig:earlystopexps} we illustrate the effect that the \mlopt bandit strategy 
has on validation error as well as on the number of total scans of the input dataset.
Models were allocated 100 iterations to converge on the correct answer. After the first 10 iterations, models that were not within 50\% of the classification error of the best model trained so far were preemptively terminated. A large percentage of models that show little or no promise of converging to a reasonable validation error are eliminated.

In the figure, the top set of bars represents the number of scans of the training data at the end of the entire search process.
The bottom set of bars represent the validation error achieved at the end of the search procedure.
The three scenarios evaluated---No Bandit, Bandit, and Baseline---represent the results of the search with no bandit allocation procedure (that is, each model is trained to completion), the algorithm the bandit allocation procedure enabled, and the baseline error rate for each dataset.

There was an 86\% decrease in total epochs across these five datasets, and the validation error is roughly comparable to the unoptimized strategy.
On average, this method achieves 93\% reduction in model error vs. not stopping
early when compared with validation error of a simple baseline model.
This relatively simple resource allocation method presents opportunities for dramatic reductions in runtime.


\subsection{Batching}
To evaluate the batching optimization, we used a synthetic dataset of $1,000,000$ data points in various dimensionality.
To illustrate the effects of amortizing scheduler overheads vs. achieving machine balance, these datasets vary in size between 750MB and 75GB.

We trained these models on a 16-node cluster of \texttt{c3.8xlarge} nodes on Amazon EC2, running Apache Spark 1.1.0.
We trained a logistic regression model on these data points via gradient descent with no batching (batch size = 1) and batching up to 20 models at once.
We implemented both a naive version of this optimization---with while loops evaluating equation~\ref{eq:grad} over each model in each task, as well as a more sophisticated version of this model which makes BLAS calls to perform the computation described in equation~\ref{eq:batchgrad}. 
For the batching experiments only, we run each algorithm for 10 iterations over the input data.
\begin{figure}
	\begin{subtable}[h]{0.4\textwidth}
	\begin{tabular}{|r|r|r|r|r|}
	  \hline
	 \diagbox{Batch Size}{D} & 100 & 1000 & 10000 \\ 
		\hline
		  1 & 826.44 & 599.60 & 553.59 \\ 
			\hline
		  2 & 1521.23 & 1214.37 & 701.07 \\ 
			\hline
		  5 & 2411.53 & 3037.97 & 992.01 \\ 
			\hline
		  8 & 5557.69 & 3502.79 & 1243.79 \\ 
			\hline
		  10 & 7148.53 & 4216.44 & 1769.12 \\
			\hline
		  15 & 7874.01 & 6260.14 & 2485.15 \\
			\hline
		  20 & 11881.18 & 8248.36 & 2445.98 \\
		   \hline
	\end{tabular}
	\caption{Models trained per hour for varying batch sizes and model complexity. Data sizes ranged from 750MB (D=100) to 75GB (D=10000).}
	\end{subtable}
	
	\begin{subtable}[h]{0.4\textwidth}
		\begin{tabular}{|r|r|r|r|r|}
		  \hline
		\diagbox{Batch Size}{D} & 100 & 1000 & 10000 \\ 
      \hline
      1 & 1.00 & 1.00  & 1.00 \\
      \hline
      2 & 1.84 & 2.02 & 1.26 \\
      \hline
      5 & 2.91 & 5.06 & 1.79 \\
      \hline
      8 & 6.72 & 5.84 & 2.24 \\
      \hline
      10 & 8.64 & 7.03 & 3.19 \\
      \hline
      15 & 9.52 & 10.44 & 4.48 \\
      \hline
      20 & 14.37 & 13.75 & 4.41 \\
      \hline
		\end{tabular}
		\caption{Speedup factor vs fastest sequential unbatched method for varying batch size and model complexity.}
	\end{subtable}
	
	\caption{Effect of batching is examined on 16 nodes with a synthetic dataset. Speedups diminish but remain significant as models increase in complexity.}
\label{fig:smallbatch}
\end{figure}

In Figure~\ref{fig:smallbatch} we show the total throughput of the system in terms of models trained per hour varying the batch size and the model complexity. 
For models trained on the smaller dataset, we see the total number of models per hour can increase by up to a factor of 15 for large batch sizes. 
This effect should not be surprising, as the actual time spent computing is on the order of milliseconds and virtually all the time goes to scheduling task execution.
In its current implementation, due to these scheduling overheads, this implementation of the algorithm under Spark will not outperform a single machine implementation for a dataset this small.
We discuss an alternative execution strategy that would better utilize cluster resources for situations where the input dataset is small in Section~\ref{sec:futureworkconclusions}.

\begin{figure}
  \centering
  \includegraphics[width=\columnwidth]{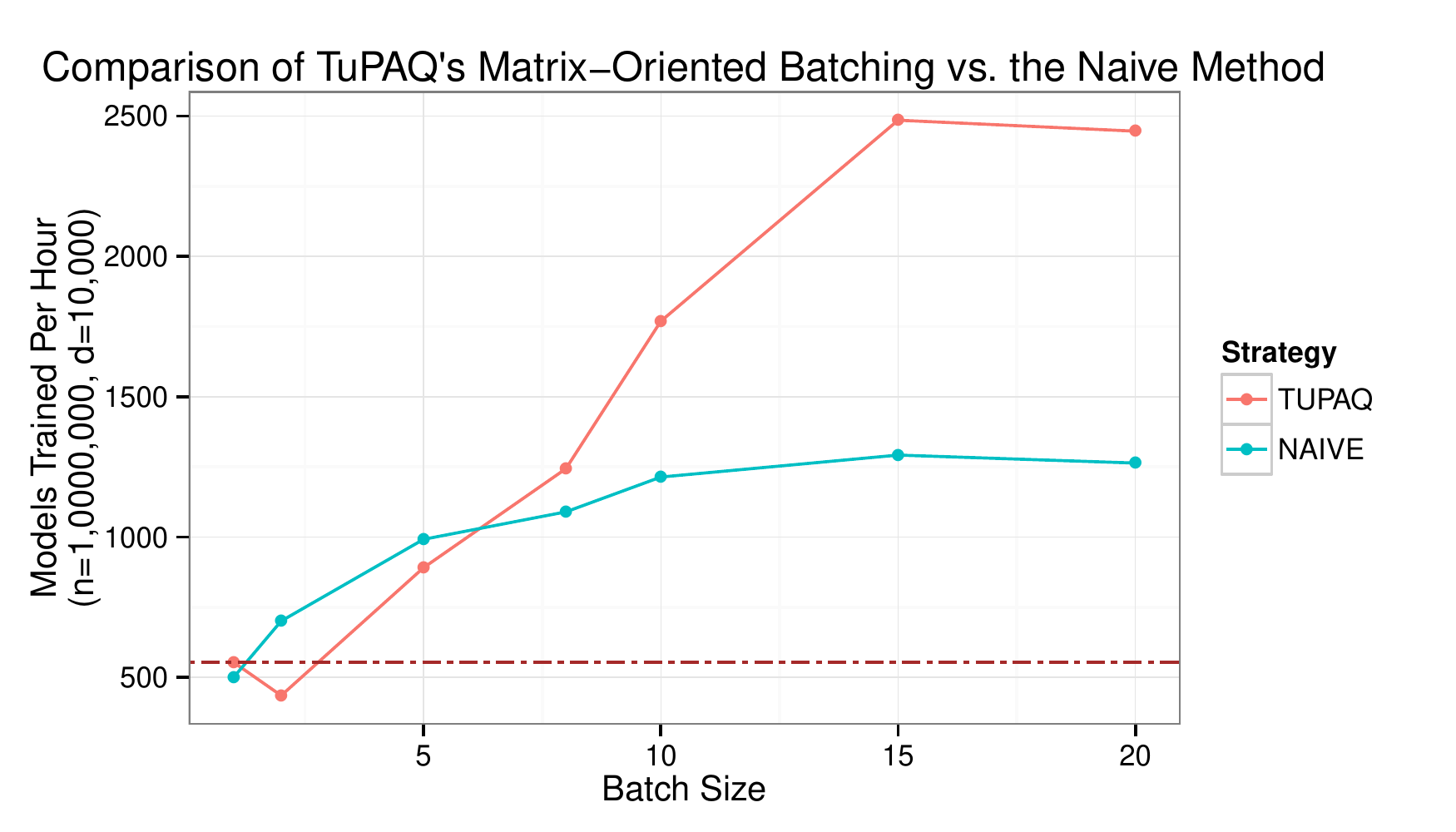}
  \caption{Leveraging high performance linear algebra libraries for batching leads to substantial speedups vs. naive methods. At bottom, we show models per hour via the fastest sequential (non-batched) strategy and demonstrate a 5x improvement in throughput.}
  \label{fig:mmbatchexps}
\end{figure}

At the other end of the spectrum in terms of data size and model complexity, we see the effects of scheduler delay start to lessen, and we maximize throughput in terms of models per hour at batch size 15.
In Figure~\ref{fig:mmbatchexps} we compare two different strategies of implementing batching---one via the naive method, and the other via the more sophisticated method---computing gradient updates via BLAS matrix multiplication.
For small batch sizes, the naive implementation actually performs faster than the BLAS optimized one.
The matrix-based implementation easily dominates the naive implementation as batch size increases. 
This is because the algorithm is slightly more cache efficient and requires only a single pass through the input data.
The overall speedup due to batching is nearly a factor of 5 when executed via matrix multiplication.

The downside to batching in the context of PAQ planning is that the system may gain information by trying plans sequentially that could inform subsequent plans that is not incorporated in later runs.
By fixing our batch size to a relatively small constant (10) we are able to balance this tradeoff.

\section{Putting It All Together}
\label{sec:evaluation}
Now that we have examined each point in the PAQ planning design space individually, let us now evaluate end-to-end performance of the \mlopt planner.

By employing batching, using state-of-the-art search methods, and using bandit resource allocation to terminate non-promising models, we are able to see a 10x increase in raw throughput of the system in terms of models trained per unit time, while finding PAQ plans that have as good or higher quality than those found with the baseline approach.

We evaluated \mlopt on very large scale data problems, at cluster sizes ranging from 16 to 128 nodes and datasets ranging from 30GB to over 3TB in size.
These sizes represent the size of the actual features the model was trained on, \emph{not} the raw data from which these features were derived.

\subsection{Platform Configuration}
We evaluated \mlopt on Linux machines running under Amazon EC2, instance type \texttt{c3.8xlarge}.
These machines were configured with Redhat Enterprise Linux, Scala 2.10, version 1.9 of the Anaconda python distribution from Continuum Analytics\cite{anaconda}, and Apache Spark 1.1.0.
Additionally, we made use of Hadoop 1.0.4 configured on local disks as our data store for the large scale experiments.
Finally, we use MLI as of commit 3e164a2d8c as a basis for \mlopt.

\subsubsection{Apache Spark Configuration}
As with any complex system, proper configuration of the platform to execute a given workload is necessary and Apache Spark is no exception.
Specifically---choosing a correct BLAS implementation, configuring spark to use it, and picking the right balance of executor threads per executor process took considerable effort. Full details of our configuration are available on request.

\subsubsection{Experimental Setup and Datasets}
The complete system involves a Scala code base built on top of Apache Spark, MLlib, and MLI.
Here, we ran experiments on 16 and 128 machines.
We used two datasets with two different learning objectives to evaluate our system at scale. 

The first dataset is a pre-featurized version of the ImageNet Large Scale Visual Recognition Challenge 2010 (ILSVRC2010) dataset~\cite{imagenet2010}, featurized using a procedure attributed to~\cite{Deng:2012tg}.
This process yields a dataset with $160,000$ features and approximately $1,200,000$ examples, or 1.4 TB of raw image features.
In our 16-node experiments we down sample to the first $16,000$ of these features and use 20\% of the base dataset for model training, which is approximately 30GB of data.
In the 128-node experiments we train on the entire dataset.
We explore five hyperparameters here---one parameter for the classifier we train---SVM or logistic regression, as well as learning rate and L2 Regularization parameters for each matching the above experiments.
We allot a budget of 128 model fittings to the problem.

As in Figure~\ref{fig:photos}, we search for a PAQ plan capable of discriminating plants from non-plants given these image features.
The images are generally in 1000 base classes, but these classes form a hierarchy and thus can be mapped into plant vs. non-plant categories.
Baseline error for this modeling task is $14.2\%$, which is a bit more skewed than the previous examples.
Our goal is to reduce validation error as much as possible, but our experience with this particular dataset has put a lower bound on validation error to around $9\%$ accuracy with linear classification models.

The second dataset is a pre-featurized version of the TIMIT Acoustic-Phonetic continuous speech corpus~\cite{timit}, featurized according to the procedure described in~\cite{timitfeatures}---yielding roughly $2,300,000$ examples each having $440$ features. While this dataset is quite small, in order to achieve strong performance on this dataset, other researchers have noted that Kernel Methods offer the best performance~\cite{posen}. Following the process of \cite{RandomKitchenSink}, this involves expanding the feature space of the dataset by nearly two orders of magnitude, yielding a dataset that has $204,800$ features. Again, this is approximately 3.4 TB of speech features.
We explore five hyperparameters here---one parameter describing the distribution family of the random projection matrix---in this case Cauchy or Gaussian, the scale and skew of these distributions, as well as the L2 regularization parameter for this model, which will have a different setting for each distribution.

A necessary precondition to supporting PAQs like those in Figure~\ref{fig:vms}, this dataset provides a examples of labeled phonemes, and our challenge is to find a model capable of labeling phonemes given some input audio. Baseline error for this modeling task is $95\%$, and state-of-the-art performance on this dataset is $35\%$ error~\cite{posen}.

\subsection{Optimization Effects}
In Figure~\ref{fig:timingexps} we can see the effects of batching and bandit allocation on the PAQ planning process for the ImageNet dataset.
Specifically, given that we want to evaluate the fitness of 128 models, it takes nearly 2 hours to fit all 128 models on the 30GB dataset of data on the 16 node cluster. 
By comparison, with the bandit rule and batching turned on, the system takes just 10 minutes to train a random search model to completion and a bit longer to train a HyperOpt model to completion, a 10x speedup in the case of random search and a 7x speedup in the case of HyperOpt.
HyperOpt takes slightly longer because it does a good job of picking points that do not need to be terminated preemptively by the bandit strategy. That is, more of the models that HyperOpt selects are trained to completion than random search. Accordingly, HyperOpt arrives at a better model than random search given the same training budget.

\begin{figure}[!htbp] \centering 
\begin{tabular}{|r|r|r|r|} 
\hline
\diagbox{Optimization}{Search Method} & Grid & Random & HyperOpt \\ 
\hline
\hline
None & $104.7$ & $100.5$ & $103.9$ \\ 
\hline
Bandits Only & $31.3$ & $29.7$ & $50.5$ \\ 
\hline
Batching Only & $31.3$ & $32.1$ & $31.8$ \\ 
\hline
All (\emph{\mlopt}) & $11.5$ & $10.4$ & $15.8$ \\ 
\hline
\end{tabular} 
\caption{Learning time in minutes for a 128-configuration budget across various optimization levels for ImageNet data. Unoptimized, sequential execution takes over 100 minutes regardless of search procedure used. Fully optimized execution can be an order of magnitude faster with \mlopt.} 
\label{fig:timingexps} 
\end{figure}

\begin{figure}
\centering
	\begin{tabular}{|r|r|r|}
	  \hline
	   \emph{Search Method} & \emph{Search Time (m)} & \emph{Test Error (\%)} \\ 
		\hline
    \hline
		  Grid (unoptimized) & 104.7 & 11.05 \\ 
			\hline
		  Random (optimized) & 10.4 & 11.41 \\ 
			\hline
		  HyperOpt (optimized) & 15.8 & 10.38 \\ 
		\hline
	\end{tabular}
  \caption{Both optimized HyperOpt and Random search perform significantly faster than unoptimized Grid search, while HyperOpt yields the best model for this image classification problem.}
\label{fig:modelconv}
\end{figure}

Turning our attention to model convergence illustrated in Figure~\ref{fig:modelconv}, we can see that on this dataset HyperOpt converges to the best answer in just 15 minutes, while random search converges to within $5\%$ of the best test error achieved by grid search a full order of magnitude faster than the baseline approach.

\subsection{Large Scale Speech and Vision}
\begin{figure}
\centering
\includegraphics[width=0.35\textwidth,scale=0.5]{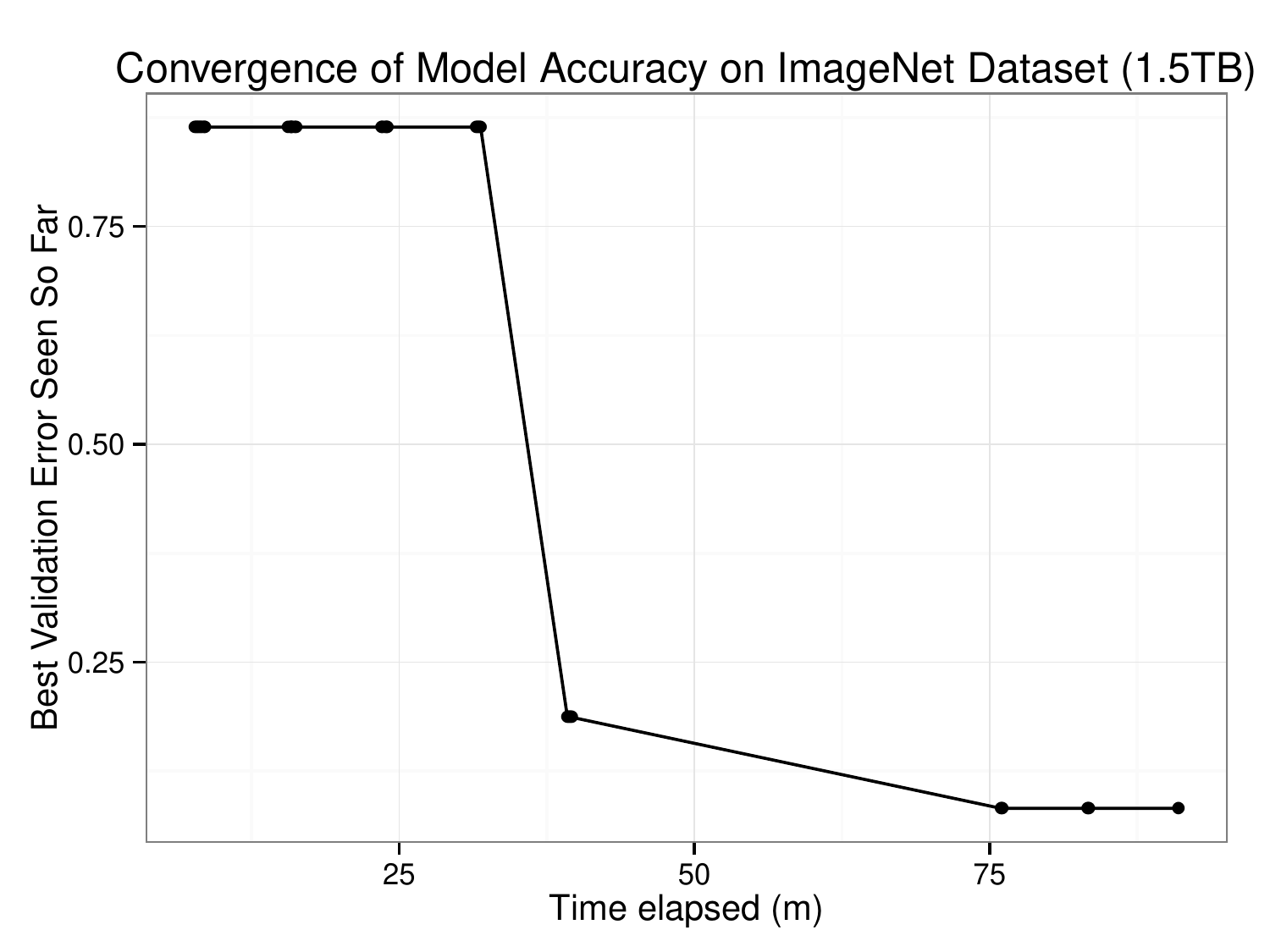}
\caption{Training a model with a budget of 32 function evaluations on a 1.2m $\times$ 160k dataset takes 90 minutes on a 128-node cluster with \mlopt.}
\label{fig:modelconvhuge}
\end{figure}

Because we employ data-parallel versions of our learning algorithms, achieving horizontal scalability with additional compute resources is trivial.
\mlopt readily scales to multi-terabyte datasets that are an order of magnitude more complicated with respect to the feature space.
For these experiments, we ran with the ImageNet models with same parameter search settings as the smaller dataset but this time with a fixed budget of 40 function evaluations.
Our results are illustrated in Figure~\ref{fig:modelconvhuge}.
Using the fully optimized HyperOpt based search method, we are able to search this space in under 90 minutes, and the method is able to achieve a validation error of $8.2\%$ for this dataset in that time.
In contrast, training all 32 models to completion using sequential grid search would have taken over 8 hours and cost upwards of \$2000.00---an expense we chose not to incur.

\begin{figure}
\centering
\includegraphics[width=0.35\textwidth,scale=0.5]{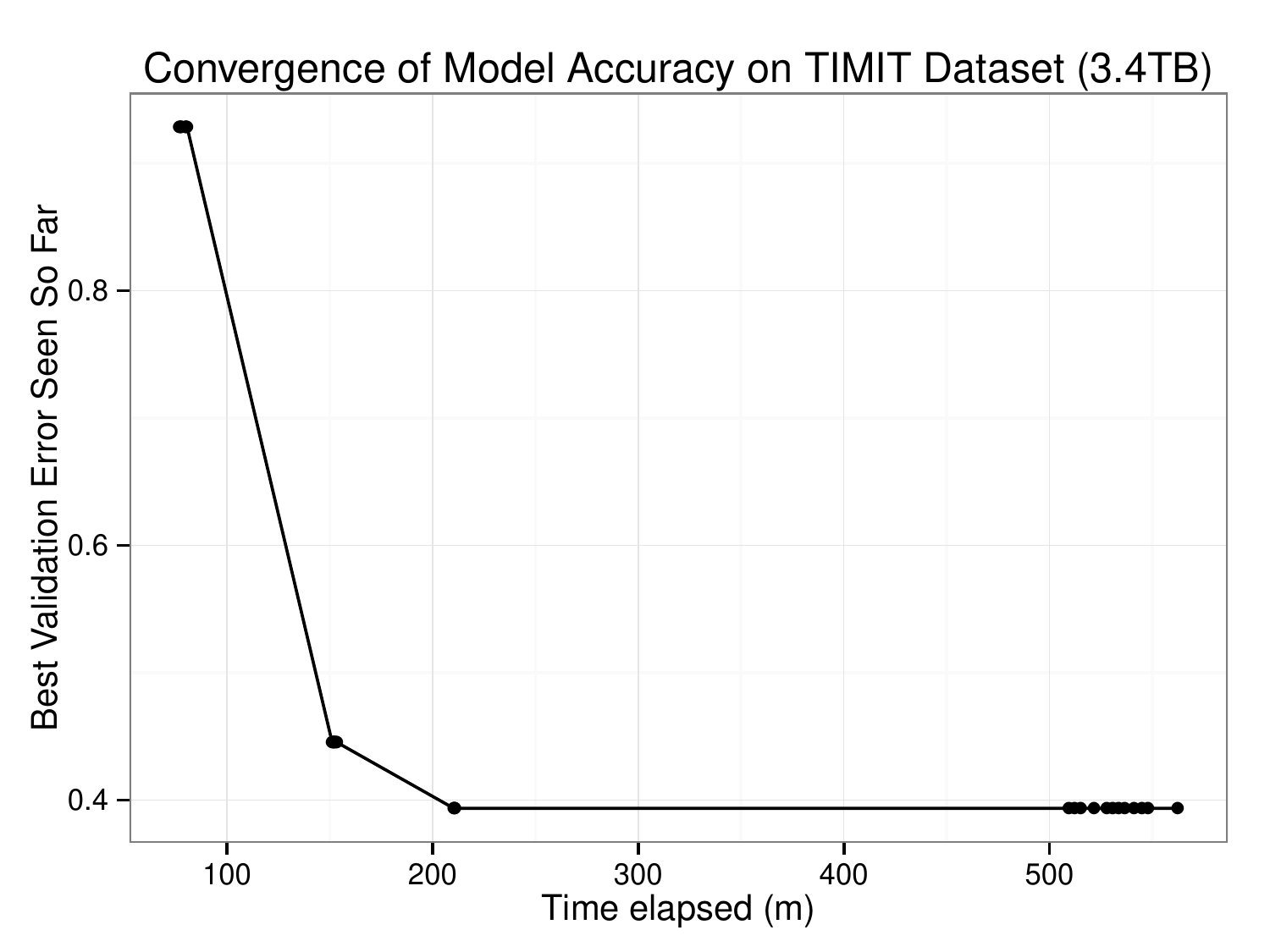}
\caption{Training a multiclass phoneme classification model on 3.7 TB of TIMIT features yields accuracy approaching that of state of the art models in 3.5 hours with \mlopt.}
\label{fig:modelconvtimit}
\end{figure}

Turning our attention to an entirely different application area, we demonstrate the ability of the system to scale to a multi-terabyte, multi-class phoneme classification problem. Here, a multi-class kernel SVM was trained on $2,251,569$ data points with $204,800$ features, in $147$ distinct classes. As shown in Figure~\ref{fig:modelconvtimit}, the system is capable of getting to a model with $39.5\%$ test error---approaching that of state-of-the-art results in speech-to-text modeling---in just 3.5 hours.
For this dataset, training the entire budget to completion would have taken 35 hours.
\section{Related Work}
\label{sec:related}
There has been a recent proliferation of systems designed for low-level,
ad-hoc distributed predictive analytics, e.g., Apache Spark~\cite{Spark},
GraphLab~\cite{powergraph}, Stratosphere~\cite{Alexandrov:2014jb}, but none
of these provide users with a declarative interface with which to specify predictive queries.

In terms of system-level optimization, both Kumar et. al.~\cite{Kumar:2013vm}
and Canny et. al.~\cite{Canny:2013th} discuss batching as an optimization for
speeding up machine learning systems.  However, \cite{Kumar:2013vm} discusses
this technique in the context of automatic feature selection, an important
problem but distinct from PAQ planning, while \cite{Canny:2013th} explores this
technique in the context of parameter exploration, model tuning, 
ensemble methods and cross validation.  We explore the impact of batching in a
distributed setting at greater depth in this work, and present a novel
application of this technique to the PAQ planning problem.

In the data mining and machine learning communities, most related to \mlopt is
Auto-WEKA \cite{Thornton:2013ea}.  As the name suggests, Auto-WEKA aims to
automate the use of Weka~\cite{weka} by applying recent derivative-free
optimization algorithms, in particular Sequential Model-based Algorithm
Configuration (SMAC)~\cite{Hutter:2011wn}, to the PAQ planning problem.  In
fact, their proposed algorithm is one of the many optimization algorithms we
use as part of \mlopt.  However, in contrast to \mlopt, Auto-WEKA focuses on
single node performance and does not optimize the parallel execution of
algorithms.  Moreover, Auto-WEKA treats algorithms as black boxes to be
executed and observed, while our system takes advantage of knowledge of
algorithm execution from both a statistical and physical perspective.

In addition to the SMAC algorithm of Auto-WEKA, other search algorithms have been recently been proposed.  In
Bergstra et. al.~\cite{Bergstra:2012ux}, the effectiveness of random search for hyperparameter
tuning is established, while Bergstra et. al.~\cite{Bergstra:2011tj} proposes a search method
that performs a random search that is refined with new
information, called Tree-structured Parzen Estimation (TPE).  We make use of both methods in our system.
Snoek et. al.~\cite{Snoek:2012vl} explore the use of Gaussian Processes for the
PAQ planning problem, and propose a variety of search algorithms, including an
algorithm that accounts for improvement per time unit, and another extension
targeting parallel implementations in which several new model configurations
are proposed at each iteration.  However, model training is nonetheless
considered a black box, and moreover, we found that their algorithms,
collectively called Spearmint, often run for several minutes per iteration even
when considering a moderate number of candidate models, which is too long to be
practical in many scenarios. 

In contrast to these recent works, the field of derivative-free optimization
has a long history of optimizing functions for which derivatives cannot be
computed~\cite{Conn:2009tl}. Our evaluation of these algorithms on the PAQ
planning problem suggests that they are not well-suited for this task,
potentially due to the lack of smoothness of the (unknown) PAQ planning
function that we are optimizing. 

There are also several proprietary and open-source systems providing machine
learning functionality with varying degrees of automation.  Google Predict
\cite{google_prediction} is Google's proprietary web-service for prediction
problems with some degree of automation, yet it restricts the maximum training
data-size to 250MB and the internals of the system are largely unknown.  

Weka~\cite{weka}, MLlib~\cite{Franklin13}, Vowpal Wabbit~\cite{vwsgd}, Hyracks~\cite{Hyracks} and
Mahout~\cite{mahout} are notable open-source ML libraries.  These systems (all distributed
with the exception of Weka),
along with proprietary projects such as SystemML~\cite{SystemML}, all focus on
training single models.
In contrast, \mlopt is designed explicitly for PAQ planning and hyperparameter
tuning at scale.  In theory our proposed methods could work with these systems,
though such integration would require these systems to
expose the access patterns of the algorithms they make available to
\mlopt.

\section{Future Work and Conclusions}
\label{sec:futureworkconclusions}

In this work, we have introduced the PAQ Planning problem, and demonstrated the
impact of logical and physical optimizations to improve the quality and
efficiency of PAQ planning.  Specifically, by combining better model search
methods, batching techniques, and bandit methods, \mlopt can find
high quality query plans for PAQs on very large datasets
an order of magnitude more efficiently than than the baseline approach.

\mlopt is a first step in tackling the challenging PAQ
planning problem.  Indeed, several avenues exist for further exploration, and
we note two broad classes of natural extensions to \mlopt.

{\bf Machine learning extensions}. From an accuracy point of view, as
additional model families are added to MLbase, \mlopt could naturally lend
itself to the construction of \emph{ensemble models} at training time -
effectively \emph{for free}.  Ensembles over a diverse set of methods are
particularly known to improve predictive performance, and there may be better
PAQ planning strategies for ensemble methods that encourage heterogeneity among
ensemble methods.  Of course, as more models and more hyperparameter
configurations are considered, PAQ planners run the risk of overfitting to the
validation data, and accounting for this issue, e.g., by \emph{controlling the
false discovery rate}~\cite{Benjamini1995}, would become especially important.
From a performance perspective, \emph{adaptive and accelerated gradient
methods} could be used to speed convergence of individual models by requiring
fewer passes over the training data~\cite{tseng08, duchi2011adaptive}.
Moreover, theoretically supported multi-armed bandit algorithms, including those
that are aware that multiple model
configurations are being sampled simultaneously, may also improve performance.
Finally, \emph{unsupervised learning methods}, including dimensionality
reduction and exploratory analysis, could be used in conjunction with
supervised approaches to speed up and/or improve the accuracy of supervised
learning methods. 

{\bf Systems extensions}. Multi-stage \emph{ML pipelines}, in which 
the initial data is transformed one or more times before being fed into 
a supervised learning algorithm, are common in most
practical ML systems.  Since each stage will likely introduce additional
hyperparameters, PAQ planning becomes more challenging 
in the pipeline setting. In a regime where a dataset is
relatively small but users still have access to cluster resources, there can be
benefits (both in terms of simplicity and speed) to broadcast the data to each
worker machine and train various models locally on each worker. PAQ planning
could be made more efficient by considering the tradeoffs between these regimes.
Training models on \emph{subsets of data} can efficiently yield
noisy evaluations of candidate models, though careful subsampling is required
to yield high-quality and trustworthy PAQ plans~\cite{BlinkDB}.
Akin to traditional query planners, PAQ
planners can learn from knowledge of the data they store and historical
workloads. A PAQ planner could store \emph{planner statistics} to tailor their
search strategy to the types of models have been used for a user's data in the
past. The evaluation of these techniques in \mlopt will be natural once the system has been exposed to a larger set of workloads.

Moving forward, we believe that \mlopt and extensions such as those
described above, will serve as a foundation for the automated construction of
end-to-end pipelines for machine learning.


\section{Acknowledgements}
This research is supported in part by NSF CISE Expeditions award CCF-1139158 and DARPA XData Award FA8750-12-2-0331, and  gifts from Amazon Web Services, Google, SAP,  Apple, Inc., Cisco, Clearstory Data, Cloudera, Ericsson, Facebook, GameOnTalis, General Electric, Hortonworks, Huawei, Intel, Microsoft, NetApp, Oracle, Samsung, Splunk, VMware, WANdisco and Yahoo!.

Thanks to Trevor Darrell, Yangqing Jia, and Sergey Karayev who provided featurized imagenet data, Ben Recht who provided valuable ideas about derivative-free optimization and feedback, and Shivaram Venkataraman, Peter Bailis, Alan Fekete, Dan Crankshaw, Sanjay Krishnan, Xinghao Pan, and Kevin Jamieson for helpful feedback.

\balance

%
\small
\bibliographystyle{abbrv}
\bibliography{refs}  
%
%
\end{document}